# Layer Tunable Third-Harmonic Generation in Multilayer Black Phosphorus

Nathan Youngblood[1], Ruoming Peng[1], Andrei Nemilentsau[1], Tony Low[1], Mo Li[1*]

[1]Department of Electrical and Computer Engineering, University of Minnesota, Minneapolis, Minnesota 55455, USA

*Corresponding Author: moli@umn.edu

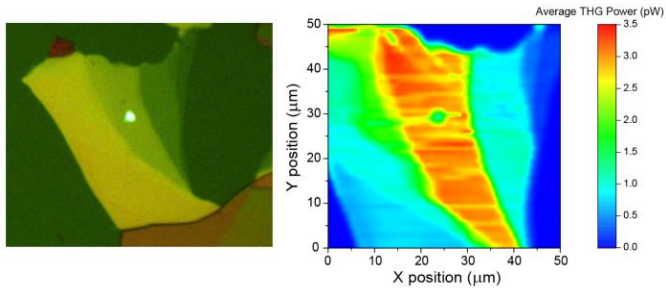

## Abstract

Black phosphorus has been the subject of growing interest due to its unique band structure that is both layer dependent and anisotropic. While many have studied the linear optical response of black phosphorus, the nonlinear response has remained relatively unexplored. Here we report on the observation of third-harmonic generation in black phosphorus using an ultrafast near-IR laser and measure $\chi^{(3)}$ experimentally for the first time. It was found that the third-harmonic emission is highly anisotropic, dependent on the incident polarization, and varies strongly with the number of layers present due to signal depletion and phase-matching conditions.

Black phosphorus (BP) recently emerged as a new two dimensional material with highly unique optical and electrical properties, including high carrier mobility,[1,2] optical and electrical anisotropy,[3–5] and importantly, a tunable, direct bandgap.[6–9] Particularly, for thicknesses greater than a few nanometers, BP's bandgap bridges the technically important mid-infrared (mid-IR) spectral range that is not currently covered by other two-dimensional semiconductors.[4] These attributes of BP make it very promising for optoelectronic devices that cover a broad spectral range from near- to mid-IR for both communication and optical sensing.[10,11]

In addition to linear optical properties, the quantum confinement in 2D materials also leads to many novel nonlinear optical properties. For example, $MoS_2$,[12–14] $WSe_2$,[15,16] and hBN,[13] all with non-centrosymmetric lattice structures, have shown strong second-order nonlinear optical effects, such as second-harmonic generation (SHG). SHG in these materials has shown strong enhancement at the exciton resonances,[15] can be electrically tuned by a local back gate,[16] and has been utilized for optically probing the crystal orientation[12] and thickness.[13,14]

Additionally, third-order optical nonlinearity such as third-harmonic generation (THG) has been observed to be strong in graphene,[17,18] as well as in $MoS_2$ thin films.[19] In terms of nonlinear optics in BP, its centrosymmetric crystalline structure only permits third-order nonlinearity but its strong anisotropy and layer dependent band structure should lead to very intriguing nonlinear optical effects. Research to date, however, has been primarily limited to the saturable absorption effect in BP, studied with z-scan[20,21] and ultrafast pump-probe[22] techniques for liquid exfoliated BP suspensions and utilized for application in mode-locked lasers.[23–27] The intrinsic optical nonlinearity of crystalline BP has not been experimentally investigated.

In this paper, we investigate BP's optical nonlinearity by measuring both the polarization and thickness dependence of THG in multilayer BP samples. We find that the THG in BP is strong and highly dependent on both the polarization of the incident light and the number of layers under investigation. From the measurement, BP's third-order nonlinear susceptibility is determined.

**Results and Discussion:**

Fig. 1a shows a typical BP flake used in this paper with the crystal orientation denoted in the lower right-hand side (x- and y-axis indicate the armchair and zigzag directions, respectively). The image shown in Fig. 1b shows the THG emission captured by the CCD camera during a measurement. Fig. 1c shows a typical spectrum of the third harmonic measured with a visible spectrometer (Ocean Optics USB4000-UV-VIS). As expected, the peak response of the THG signal (519 nm) is exactly one third of the fundamental wavelength of the pump laser (1557 nm). Further confirmation of THG can be seen in Fig. 1d, which plots the pump power dependence of the THG power with a power law fit to the experimental data.

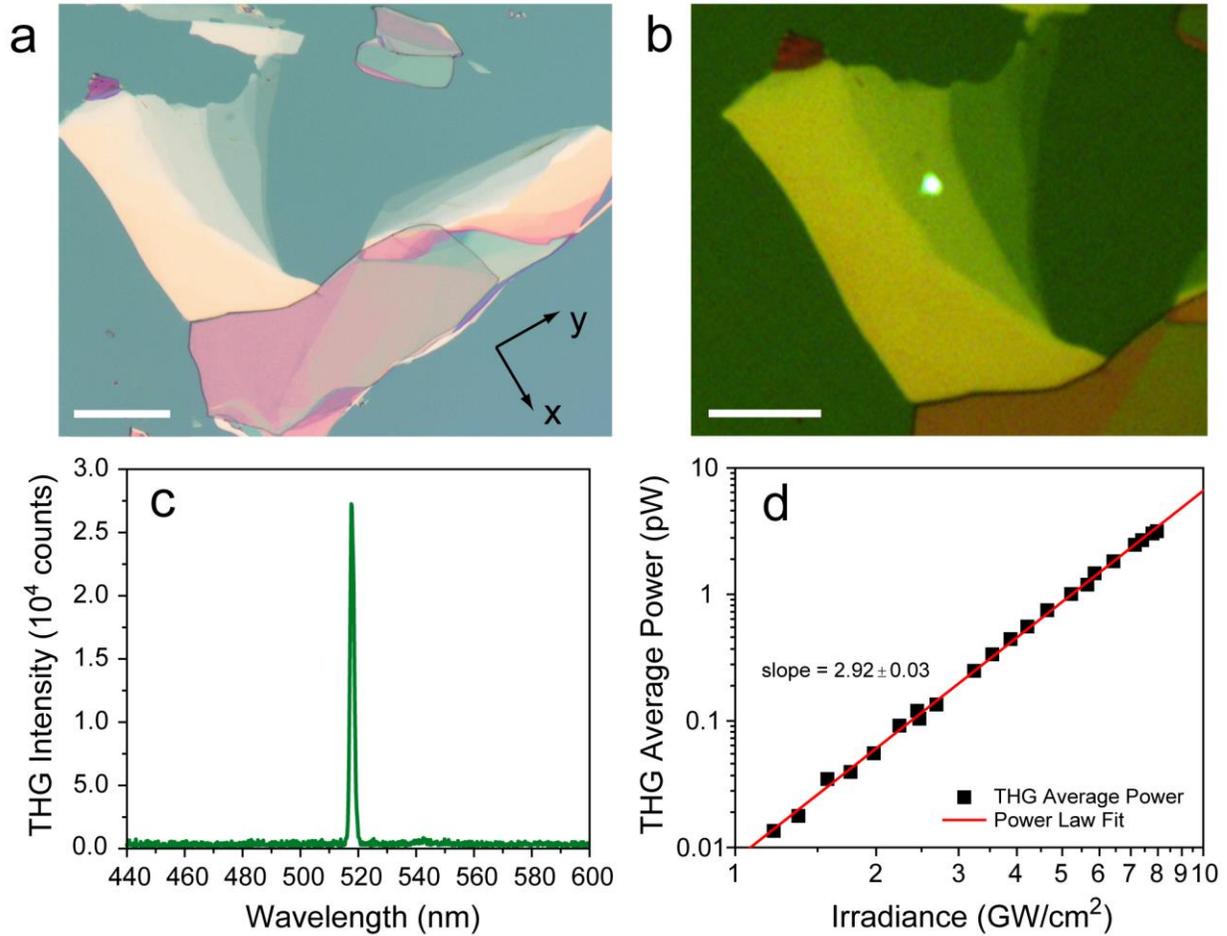

**Figure 1:** Third harmonic generation (THG) in multilayer BP. (a) Optical image of black phosphorus flake with multiple layers visible (scale bar 25 μm). Crystal orientation indicated in lower right. (b) Third harmonic emission (bright spot) from the flake in (a) measured by a CCD camera (scale bar 20 μm). (c) Measured spectrum of THG emission with a peak wavelength at 519 nm, which is three times the frequency of the fundamental excitation. (d) The power dependence of THG from BP is plotted with a power law fit to the data.

Since the linear optical response of BP shows strong anisotropy with the incident light's polarization,[3,28,29] it can be expected that the nonlinear response would show similar anisotropy. The blue triangles in Fig. 2a show the angular dependence of the THG emission as a function of incident polarization relative to the x-axis (armchair direction) of the crystal. The minimum in the THG corresponds to an incident polarization aligned along the y-axis (zigzag direction) of the crystal. This was verified by photocurrent measurements on other BP flakes with a similar thickness. The polarization angle that minimized the photocurrent (as seen in Yuan *et al.*[30]) also minimized the emitted THG power. To extract the polarization dependence of the THG emission, an analyzer was placed in front of the APD and fixed at either 0 or 90 degrees while a near-IR, half-wave plate was rotated in 5-degree steps to change the linear polarization of the pump laser. The black and red data points in Fig. 2a and b correspond to the measured intensity of the *x* and *y* polarized light, respectively.

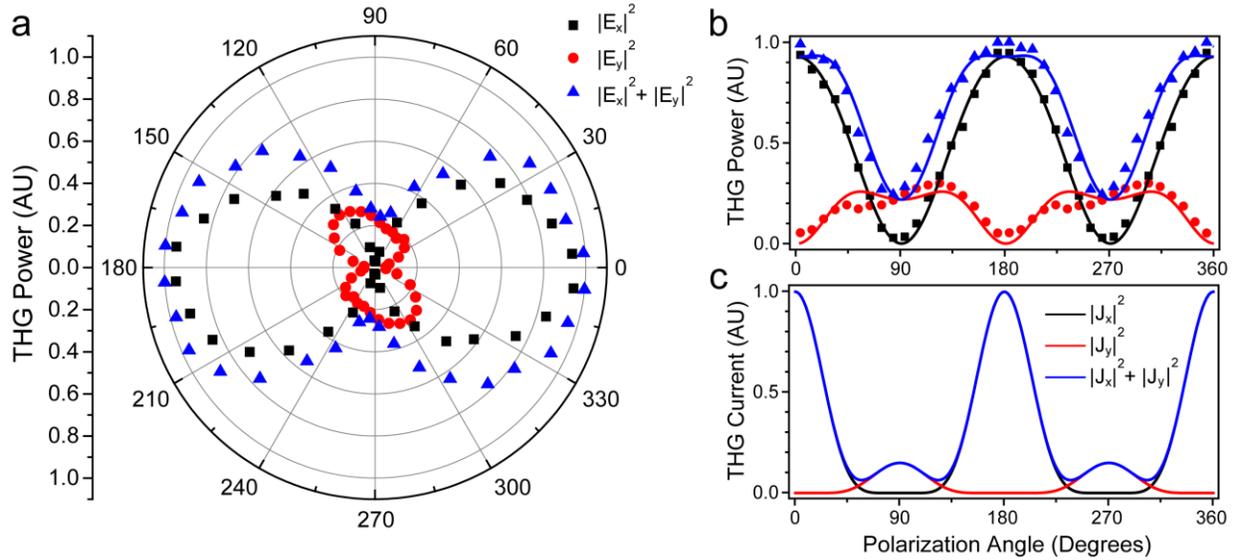

**Figure 2:** Anisotropic THG in BP. (a) Dependence of THG on the incident polarization. Zero degrees corresponds to the x-axis (armchair direction) of the BP crystal. (b) Same data as in (a) with theoretical fits to the intensity polarized in the *x* (black squares) and *y* (red dots) directions. Blue triangles correspond to the total intensity. (c) Calculated THG current induced in BP.

BP has an orthorhombic crystal structure and is in the space group Cmca.[31] Knowing this, we can write the contracted third-order nonlinear susceptibility tensor for an orthorhombic crystal exhibiting THG as follows:[32]

$$\begin{bmatrix} \chi_{11} & 0 & 0 & 0 & 0 & \chi_{16} & 0 & \chi_{18} & 0 & 0 \\ 0 & \chi_{22} & 0 & \chi_{24} & 0 & 0 & 0 & 0 & \chi_{29} & 0 \\ 0 & 0 & \chi_{33} & 0 & \chi_{35} & 0 & \chi_{37} & 0 & 0 & 0 \end{bmatrix} \quad (1)$$

where the first subscript "1, 2, 3" refer to "x, y, z" respectively and the second subscript signifies the following:

| jkl | xxx | yyy | zzz | yzz | yyz | xzz | xxz | xyy | xxy | xyz |
|---|---|---|---|---|---|---|---|---|---|---|
| m | 1 | 2 | 3 | 4 | 5 | 6 | 7 | 8 | 9 | 0 |

Since we are only exciting in the x-y plane in our configuration, we can set all components that contain a z term to zero. We are left with only four non-zero matrix elements $(\chi_{11}, \chi_{22}, \chi_{18}, \chi_{29})$, which determine the THG in BP. Because the third-harmonic electric field is proportional to the nonlinear susceptibility and we can write the THG output intensity as follows:

$$\begin{aligned} E_x^2 &\propto \left[ \chi_{11} \cos^3(\theta) + \chi_{18} \cos(\theta)\sin^2(\theta) \right]^2 \\ E_y^2 &\propto \left[ \chi_{22} \sin^3(\theta) + \chi_{29} \sin(\theta)\cos^2(\theta) \right]^2 \end{aligned} \quad (2)$$

where $\theta$ is the polarization angle relative to the x-axis of the crystal. The solid lines in Fig. 2b are fits to the measured intensity in the *x* and *y* directions using the relationships in Equation 2. One can see good agreement with the overall fit, excluding some asymmetry in the $|E_y|^2$ data. This asymmetry is likely caused by some residual polarization-dependence in the transmission of the dichroic mirror that we were not able to account for during the measurement.

In order to confirm our interpretation, we calculated the density, $j^{(3)}(3\omega)$, of the nonlinear electric current induced in BP at the frequency of third harmonic by the pump field, $E_p = (1/2)E_0 e^{-i\omega t} + c.c.$. We studied dynamics of the electron subsystem by solving the Liouville equation for a single-electron density operator, $i\hbar \partial_t \rho = [H, \rho]$, where $H=H_0+H_{int}+H_{rel}$ is a single particle Hamiltonian, $H_0$ is a Hamiltonian of the unperturbed electron system, $H_{int}$ describes the interaction between an electron and the pump, and $H_{rel}$ is the relaxation term (see supporting information and Keller et al.[33] and Nemilentsau et al.[34] for details). We solved the Liouville equation in the third order of perturbation theory, $\rho = \sum_{n=0}^{3} \rho^{(3)}$, where $\rho^{(3)} = (1/2)\rho^{(3;3\omega)}e^{-3i\omega t} + (1/2)\rho^{(3;\omega)}e^{-i\omega t} + c.c.$. The term $\rho^{(3;3\omega)}$ describes THG, while $\rho^{(3;\omega)}$ accounts for the self-action effects. The density of the nonlinear current is calculated as $\boldsymbol{j}^{(3)}(3\omega) = \text{Tr}\left[\rho^{(3;3\omega)}\boldsymbol{J}\right]$, where $\boldsymbol{J}$ is the current density operator.

In order to keep the calculations tractable, we consider the case of monolayer BP. In this case, the low energy Hamiltonian is available[29] (see also Supporting Information). The simulation results presented in Fig. 2c considers the situation where the pump field is at 0.5 eV, such that the three photon processes are in resonance with the monolayer gap of 1.5 eV. As one can see, the numerical results are in good qualitative agreement with experimental data. The values of induced current are maximum when pump field is polarized along x-direction, and minimum when polarization is along y direction. The strong anisotropy is direct consequence of the band parameters anisotropy, with the effective mass $m_{xx}^*$ component being considerably smaller than $m_{yy}^*$. Although the experimental conditions (i.e. frequency and layer numbers) differs from our calculation, we expect similar angular dependence of the induced current at higher frequencies as well, since strong anisotropy of the energy bands imparted by the symmetry of the BP crystalline lattice is also expected far from the band edge.[35]

To further characterize the nature of THG in BP, we recorded the power of the THG signal as a function of position by scanning the sample in an x-y grid of 500 nm increments. This allowed us to create a two-dimensional map of the THG power as can be seen in Fig. 3b. We also used atomic force microscopy (AFM) to measure the topography of the various layers present in the same region (Fig. 3a). A few wrinkles can be seen in the AFM image that are due to the PDMS transfer process. Interestingly, the two larger wrinkles at the top of the flake can also be seen in the THG map of Fig. 3b. Modulation in the optical properties of wrinkled BP has recently been studied where the strain due to wrinkles strongly modifies the band structure and therefore the optical properties of BP.[36]

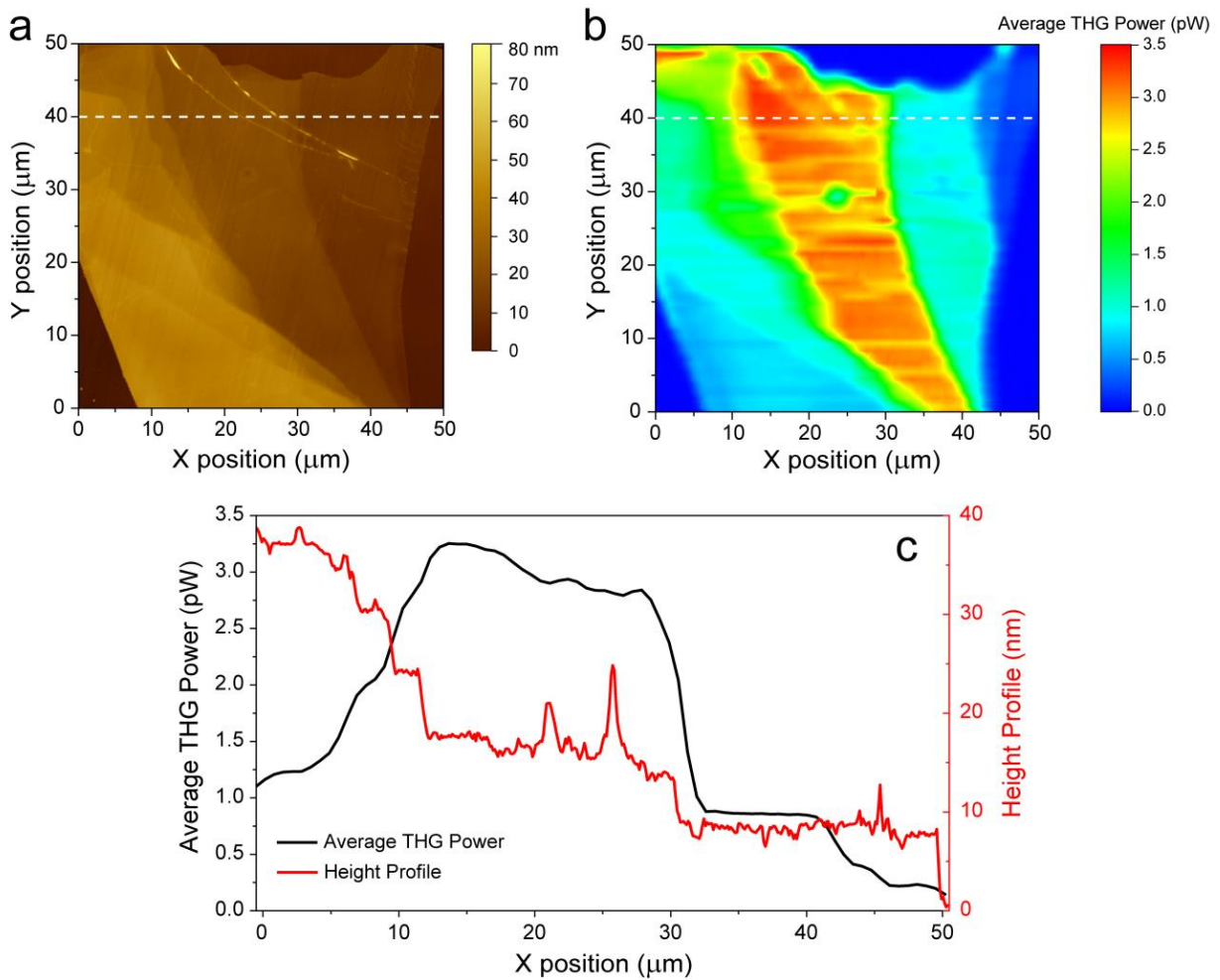

**Figure 3:** Mapping of THG signal. (a) Atomic force microscopy (AFM) image of a flake of BP containing many layers of various thicknesses. (b) Position dependent THG signal measured by scanning the position of the beam relative to the sample. Different thicknesses can be resolved *via* contrast in the THG signal. (c) Profiles of both the height and THG power extracted from the white cut-lines shown in (a) and (b). THG can show higher contrast than AFM for flake thicknesses less than 15 nm.

Fig. 3c shows the cut-lines for both the topography and THG images denoted by the white lines in Fig. 3a and b. One can see that the THG varies significantly as a function of position, but is relatively constant for a fixed flake thickness. Additionally, for thicknesses less than 15 nm, there is much higher contrast between various thicknesses in the THG image compared to the AFM image. We attribute this significant change in the THG power to be largely due to depletion of the THG signal and phase mismatch between the fundamental and THG.

By measuring the average THG power at different positions from Fig. 3b and relating it to the height from AFM data in Fig. 3a, the THG power as a function of thickness can be extracted. Fig. 4a plots the average THG power verses BP thickness from multiple samples. This thickness dependence can be primarily understood by two competing mechanisms. The first is a quadratic increase in the THG signal that is proportional to the square of the number of layers. This assumption is justified for our flakes in the range of 6 to 15 nm where the thickness is less than the coherence length[37,38] (backward propagating wave: $L_{coh} = \lambda/6(n_\omega + n_{3\omega}) \approx 40$ nm and forward propagating wave: $L_{coh} = \lambda/6(n_\omega - n_{3\omega}) \approx 870$ nm). As the thickness increases, however, the strong absorption of BP at visible wavelengths contributes to significant depletion of the THG signal as it propagates up through the flake. This can be observed in the exponential decay of the signal at thicknesses greater than 15 nm. Fitting the experimental data with an exponential decay, we estimate the extinction coefficient (the imaginary part of the refractive index) to be $\kappa \approx 2.5$. This value is a factor of 2 larger than the value reported by Mao *et al.* for BP in the visible regime.[39] In order to fully account for the depletion of the THG and interference effects of both the fundamental and third-harmonic signal, we performed FDTD simulations using Lumerical Solutions. The excitation source used in the simulation has the same pulse energy and bandwidth as the ultrafast laser used in our experiments. The simulation results are compared with our experimental measurements in Fig. 4a, showing a good agreement. However, at small BP thicknesses (< 10 nm), the simulation shows the same trend as the experiment, but the absolute values deviate as the flake thickness decreases. One possible explanation is that the exfoliation, transfer, and $Al_2O_3$ growth negatively affect the first few layers which accounts for a larger fraction of the total flake for thinner flakes, thus reducing the THG yield. Further studies using fabrication in a more inert environment and van der Waals passivation to preserve the pristine conditions of BP could shed light on this discrepancy.

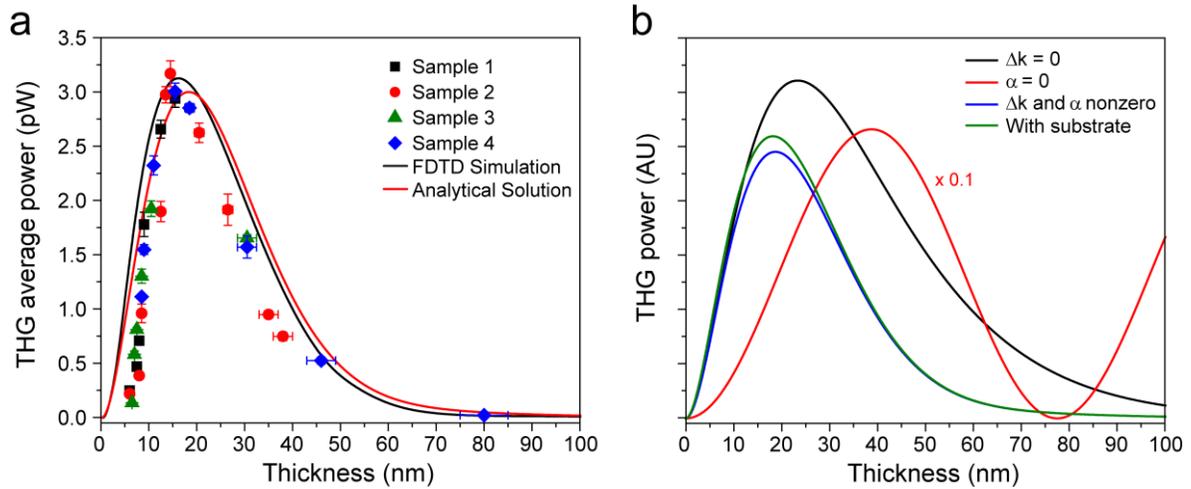

**Figure 4:** THG dependence on BP thickness. (a) THG average power plotted verses thickness (symbols). Data was extracted from the AFM and THG map of four BP flakes. The solid black line corresponds to full FDTD simulation where $\chi^{(3)}$ was assumed to be constant and the thickness of the BP layer was varied on a 300 nm $SiO_2$/Si substrate. The solid red line corresponds to the analytical solution which includes absorption of the THG signal which is derived in the supporting information. (b) Analytical model of THG in BP showing the contributions of optical absorption, phase mismatch, and reflection of THG from substrate.

In order to intuitively understand the general trend in Fig. 4a, we plot the thickness dependent portion of the analytical solution for THG by solving nonlinear Maxwell's equations (see section 2 of Supporting Information):

$$I_{3\omega}(d) = A(\omega)|\chi^{(3)}|^2 I_\omega^3 \left(\frac{e^{-2\alpha d} - 2\cos(\Delta k d)\, e^{-\alpha d} + 1}{\alpha^2 + \Delta k^2}\right) e^{-2\alpha d}$$

where $d$ is the thickness of the BP flake, $\alpha$ is the absorption coefficient of the THG signal, and $A(\omega)$ is a scaling factor which includes physical constants and the refractive indices at $\omega$ and $3\omega$. The results of this analytical model can be seen in Fig. 4b where we compare the various effects that optical absorption, phase mismatch, and reflection from substrate have on the general shape of the curve. We note that although this analytical solution does not consider the absorption and interference of the fundamental pump, simply taking into account the phase mismatch and the absorption of the third-harmonic is enough to describe the general trend we observe in the data. The full FDTD simulations we performed in Fig. 4a account for these secondary effects.

It is possible to extract the magnitude of $\chi^{(3)}$ at the peak response (14.5 nm thick) using the result for a thin bulk sample with loss that we derive in the Supporting Information. Taking into account the average power, repetition rate, and spot size of the ultrafast laser (30 mW, 35 MHz, and 5.7 μm diameter at FWHM) and the refractive index of BP[29,39] ($n_\omega \approx 3.2 + 0.2i$, $n_{3\omega} \approx 3.5 + 2.5i$), we estimate the magnitude of $\chi^{(3)}$ to be $(1.4 \pm 0.1) \times 10^{-19}$ m²/V².

Table 1 compares our results to the experimentally measured values of $\chi^{(3)}$ in other 2D materials reported in the literature. Something directly evident from Table 1 is the wide range of reported results for $\chi^{(3)}$ in graphene. This also influences the estimated $\chi^{(3)}$ for other materials where the authors directly compare the magnitude of THG in MoS$_2$ and GaSe with a graphene flake on the same substrate and scale $\chi^{(3)}$ accordingly. One explanation is the effect the substrate has on the magnitude of the THG signal observed.[43] Constructive interference of the fundamental pump due to reflections from the substrate can lead to significant enhancements in the power at the graphene layer and lead to overestimates of $\chi^{(3)}$ since the THG signal is proportional to $I_1^3$. In our FDTD simulations, we found that using $\chi^{(3)} = 10^{-19}$ m²/V² gave a very comparable THG power for the same pulse energy and bandwidth that we used in our experiments so we believe our estimate of $\chi^{(3)} \sim 1.4 \times 10^{-19}$ m²/V² from the analytical model to be quite close to the actual value. Another explanation, as pointed out in Woodward et al.,[40] is a likely miscalculation in the original estimation of graphene's $\chi^{(3)}$ by Hendry et al.,[44] which leads to a corrected value of $\chi^{(3)} \sim 10^{-19}$ m²/V².[45]

**Table 1. List of $\chi^{(3)}$ values of various 2D materials reported in the literature**

| Material | Number of Layers | $\chi^{(3)}$ (m²/V²) | Fundamental Wavelength (nm) | Reference |
|---|---|---|---|---|
| BP | 29 | $1.4 \times 10^{-19}$ | 1557 | this work |
| MoS$_2$ | 1 | $2.9 \times 10^{-19}$ | 1560 | [40] |
| MoS$_2$ | >7 | $\sim 10^{-19}$ | 1758 | [19] |
| MoS$_2$ | 1 to 7 | $3.9 \times 10^{-15}$ | 1560 | [41] |
| GaSe | 9 to 14 | $1.7 \times 10^{-16}$ | 1560 | [38] |
| Graphene | 1 | $\sim 10^{-16}$ | 1720 | [17] |
| Graphene | 1 | $4 \times 10^{-15}$ | 1550 | [42] |
| Graphene | 1 | $1.5 \times 10^{-19}$ | 1560 | [40] |
| Graphene | 1 and 2 | $\sim 10^{-19}$ | 789 | [18] |

Our estimate of $\chi^{(3)}$ is typical for other bulk materials that exhibit a strong third-order optical response[37] but an order of magnitude lower than our theoretical prediction of $\chi^{(3)}$ at the band edge (see Supporting Information). It is indeed rather intriguing that the magnitude of $\chi^{(3)}$ in BP is comparable to other 2D materials when the excitation energy used is so far above the bandgap of BP. While our initial results are promising, it will be extremely interesting in future works to characterize $\chi^{(3)}$ near the band edge in BP of different thicknesses, which will require a pump photon energy in the mid-IR regime that is not available in our measurement system.

**Conclusions:**

In conclusion, we have observed THG in BP thin films using an ultrafast near-IR laser. The third-order response was found to be highly anisotropic with incident polarization. Additionally, we found that the THG was strongly dependent on thickness due to absorption of the third-harmonic signal by BP in addition to interference effects introduced by the substrate. Our study shows that multilayer BP's strong nonlinear and anisotropic optical response could be useful for nonlinear applications. Particularly, at the band edge of BP, that is for mid-infrared band light, BP's nonlinearity is expected to be very strong and therefore promising for application in mid-infrared nonlinear optics.

**Note:** During the review process we became aware of another work that studied the third-harmonic response of laser thinned flakes of BP.[46]

**Methods:**

To prepare our samples, we used Scotch tape to exfoliate single-crystal BP onto glass slides covered with polydimethylsiloxane (PDMS). Once on the PDMS, suitable BP flakes with multiple layers were identified under an optical microscope and transferred onto silicon substrates coated with 300 nm of thermally grown $SiO_2$. After transferring the BP, the substrates were immediately placed in an atomic layer deposition (ALD) chamber and coated with 20 nm $Al_2O_3$ grown at 180 C for passivation to prevent oxidation of the material.

A femtosecond, near-IR fiber laser with a spectrum centered at 1557 nm (photon energy ~0.8 eV) and repetition rate of 35 MHz was used as the pump (200 fs pulse width and 4.4 kW peak power) to probe the THG response of our samples. Light from the laser was focused onto the sample *via* a near-IR 20× objective (Mitutoyo MPlan NIR, 0.4 NA). A short-pass dichroic filter was used to reflect the fundamental frequency toward the sample, while passing the third harmonic. The filtered light was focused onto either a CCD camera, spectrometer, or avalanche photodetector (APD) depending on the measurement.

**Supporting Information:**

The Supporting Information is available free of charge on the [ACS Publications website](#) at DOI: XXX

- Theory of modeling third-harmonic generation; THG-depletion model of THG verses thickness; THG maps and optical images of additional samples used to create Figure 4; Third-order autocorrelation measurement

# References:

(1) Li, L.; Yu, Y.; Ye, G. J.; Ge, Q.; Ou, X.; Wu, H.; Feng, D.; Chen, X. H.; Zhang, Y. Black Phosphorus Field-Effect Transistors. *Nat. Nanotechnol.* **2014**, *9*, 372–377 DOI:10.1038/nnano.2014.35.

(2) Liu, H.; Neal, A. T.; Zhu, Z.; Luo, Z.; Xu, X.; Tománek, D.; Ye, P. D. Phosphorene: An Unexplored 2D Semiconductor with a High Hole Mobility. *ACS Nano* **2014**, *8*, 4033–4041 DOI:10.1021/nn501226z.

(3) Qiao, J.; Kong, X.; Hu, Z.-X.; Yang, F.; Ji, W. High-Mobility Transport Anisotropy and Linear Dichroism in Few-Layer Black Phosphorus. *Nat. Commun.* **2014**, *5*, 4475 DOI:10.1038/ncomms5475.

(4) Xia, F.; Wang, H.; Xiao, D.; Dubey, M.; Ramasubramaniam, A. Two-Dimensional Material Nanophotonics. *Nat. Photonics* **2014**, *8*, 899–907 DOI:10.1038/nphoton.2014.271.

(5) Wang, X.; Jones, A. M.; Seyler, K. L.; Tran, V.; Jia, Y.; Zhao, H.; Wang, H.; Yang, L.; Xu, X.; Xia, F. Highly Anisotropic and Robust Excitons in Monolayer Black Phosphorus. *Nat. Nanotechnol.* **2015**, *10*, 517–521 DOI:10.1038/nnano.2015.71.

(6) Takao, Y.; Asahina, H.; Morita, A. Electronic Structure of Black Phosphorus in Tight Binding Approach. *J. Phys. Soc. Japan* **1981**, *105*, 3362–3369 DOI:10.1143/JPSJ.50.3362.

(7) Tran, V.; Soklaski, R.; Liang, Y.; Yang, L. Layer-Controlled Band Gap and Anisotropic Excitons in Few-Layer Black Phosphorus. *Phys. Rev. B* **2014**, *89*, 235319 DOI:10.1103/PhysRevB.89.235319.

(8) Kim, J.; Baik, S. S.; Ryu, S. H.; Sohn, Y.; Park, S.; Park, B.-G.; Denlinger, J.; Yi, Y.; Choi, H. J.; Kim, K. S. Observation of Tunable Band Gap and Anisotropic Dirac Semimetal State in Black Phosphorus. *Science* **2015**, *349*, 723–726 DOI:10.1126/science.aaa6486.

(9) Das, S.; Zhang, W.; Demarteau, M.; Hoffmann, A.; Dubey, M.; Roelofs, A. Tunable Transport Gap in Phosphorene. *Nano Lett.* **2014**, *14*, 5733–5739 DOI:10.1021/nl5025535.

(10) Youngblood, N.; Chen, C.; Koester, S. J.; Li, M. Waveguide-Integrated Black Phosphorus Photodetector with High Responsivity and Low Dark Current. *Nat. Photonics* **2015**, *9*,

249–252 DOI:10.1038/nphoton.2015.23.

(11) Lin, C.; Grassi, R.; Low, T.; Helmy, A. S. Multilayer Black Phosphorus as a Versatile Mid-Infrared Electro-Optic Material. *Nano Lett.* **2016**, *16*, 1683–1689 DOI:10.1021/acs.nanolett.5b04594.

(12) Yin, X.; Ye, Z.; Chenet, D. A.; Ye, Y.; O'Brien, K.; Hone, J. C.; Zhang, X. Edge Nonlinear Optics on a MoS2 Atomic Monolayer. *Science* **2014**, *344*, 488–490 DOI:10.1126/science.1250564.

(13) Li, Y.; Rao, Y.; Mak, K. F.; You, Y.; Wang, S.; Dean, C. R.; Heinz, T. F. Probing Symmetry Properties of Few-Layer MoS2 and H-BN by Optical Second-Harmonic Generation. *Nano Lett.* **2013**, *13*, 3329–3333 DOI:10.1021/nl401561r.

(14) Malard, L. M.; Alencar, T. V.; Barboza, A. P. M.; Mak, K. F.; de Paula, A. M. Observation of Intense Second Harmonic Generation from MoS2 Atomic Crystals. *Phys. Rev. B* **2013**, *87*, 201401 DOI:10.1103/PhysRevB.87.201401.

(15) Wang, G.; Marie, X.; Gerber, I.; Amand, T.; Lagarde, D.; Bouet, L.; Vidal, M.; Balocchi, A.; Urbaszek, B. Giant Enhancement of the Optical Second-Harmonic Emission of WSe(2) Monolayers by Laser Excitation at Exciton Resonances. *Phys. Rev. Lett.* **2015**, *114*, 97403 DOI:10.1103/PhysRevLett.114.097403.

(16) Seyler, K. L.; Schaibley, J. R.; Gong, P.; Rivera, P.; Jones, A. M.; Wu, S.; Yan, J.; Mandrus, D. G.; Yao, W.; Xu, X. Electrical Control of Second-Harmonic Generation in a WSe2 Monolayer Transistor. *Nat. Nanotechnol.* **2015**, *10*, 407–411 DOI:10.1038/nnano.2015.73.

(17) Kumar, N.; Kumar, J.; Gerstenkorn, C.; Wang, R.; Chiu, H.-Y.; Smirl, A. L.; Zhao, H. Third Harmonic Generation in Graphene and Few-Layer Graphite Films. *Phys. Rev. B* **2013**, *87*, 121406 DOI:10.1103/PhysRevB.87.121406.

(18) Hong, S.-Y.; Dadap, J. I.; Petrone, N.; Yeh, P.-C.; Hone, J.; Osgood, R. M. Optical Third-Harmonic Generation in Graphene. *Phys. Rev. X* **2013**, *3*, 21014 DOI:10.1103/PhysRevX.3.021014.

(19) Wang, R.; Chien, H.-C.; Kumar, J.; Kumar, N.; Chiu, H.-Y.; Zhao, H. Third-Harmonic Generation in Ultrathin Films of MoS2. *ACS Appl. Mater. Interfaces* **2014**, *6*, 314–318 DOI:10.1021/am4042542.

(20) Lu, S. B.; Miao, L. L.; Guo, Z. N.; Qi, X.; Zhao, C. J.; Zhang, H.; Wen, S. C.; Tang, D.

Y.; Fan, D. Y. Broadband Nonlinear Optical Response in Multi-Layer Black Phosphorus: An Emerging Infrared and Mid-Infrared Optical Material. *Opt. Express* **2015**, *23*, 11183–11194 DOI:10.1364/OE.23.011183.

(21) Zhang, F.; Wu, Z.; Wang, Z.; Wang, D.; Wang, S.; Xu, X. Strong Optical Limiting Behavior Discovered in Black Phosphorus. *RSC Adv.* **2016**, *6*, 20027–20033 DOI:10.1039/C6RA01607C.

(22) Wang, Y.; Huang, G.; Mu, H.; Lin, S.; Chen, J.; Xiao, S.; Bao, Q.; He, J. Ultrafast Recovery Time and Broadband Saturable Absorption Properties of Black Phosphorus Suspension. *Appl. Phys. Lett.* **2015**, *107*, 91905 DOI:10.1063/1.4930077.

(23) Sotor, J.; Sobon, G.; Macherzynski, W.; Paletko, P.; Abramski, K. M. Black Phosphorus Saturable Absorber for Ultrashort Pulse Generation. *Appl. Phys. Lett.* **2015**, *107*, 51108 DOI:10.1063/1.4927673.

(24) Luo, Z.-C.; Liu, M.; Guo, Z.-N.; Jiang, X.-F.; Luo, A.-P.; Zhao, C.-J.; Yu, X.-F.; Xu, W.-C.; Zhang, H. Microfiber-Based Few-Layer Black Phosphorus Saturable Absorber for Ultra-Fast Fiber Laser. *Opt. Express* **2015**, *23*, 20030 DOI:10.1364/OE.23.020030.

(25) Mu, H.; Lin, S.; Wang, Z.; Xiao, S.; Li, P.; Chen, Y.; Zhang, H.; Bao, H.; Lau, S. P.; Pan, C.; Fan, D.; Bao, Q. Black Phosphorus-Polymer Composites for Pulsed Lasers. *Adv. Opt. Mater.* **2015**, *3*, 1447–1453 DOI:10.1002/adom.201500336.

(26) Chen, Y.; Jiang, G.; Chen, S.; Guo, Z.; Yu, X.; Zhao, C.; Zhang, H.; Bao, Q.; Wen, S.; Tang, D.; Fan, D. Mechanically Exfoliated Black Phosphorus as a New Saturable Absorber for Both Q-Switching and Mode-Locking Laser Operation. *Opt. Express* **2015**, *23*, 12823–12833 DOI:10.1364/OE.23.012823.

(27) Lu, D.; Pan, Z.; Zhang, R.; Xu, T.; Yang, R.; Yang, B.; Liu, Z.; Yu, H.; Zhang, H.; Wang, J. Passively Q-Switched Ytterbium-Doped ScBO3 Laser with Black Phosphorus Saturable Absorber. *Opt. Eng.* **2016**, *55*, 81312 DOI:10.1117/1.OE.55.8.081312.

(28) Xia, F.; Wang, H.; Jia, Y. Rediscovering Black Phosphorus as an Anisotropic Layered Material for Optoelectronics and Electronics. *Nat. Commun.* **2014**, *5*, 4458 DOI:10.1038/ncomms5458.

(29) Low, T.; Rodin, A. S.; Carvalho, A.; Jiang, Y.; Wang, H.; Xia, F.; Castro Neto, A. H. Tunable Optical Properties of Multilayer Black Phosphorus Thin Films. *Phys. Rev. B* **2014**, *90*, 75434 DOI:10.1103/PhysRevB.90.075434.

(30) Yuan, H.; Liu, X.; Afshinmanesh, F.; Li, W.; Xu, G.; Sun, J.; Lian, B.; Curto, A. G.; Ye, G.; Hikita, Y.; Shen, Z.; Zhang, S.-C.; Chen, X.; Brongersma, M.; Hwang, H. Y.; Cui, Y. Polarization-Sensitive Broadband Photodetector Using a Black Phosphorus Vertical P-N Junction. *Nat. Nanotechnol.* **2015**, *10*, 707–713 DOI:10.1038/nnano.2015.112.

(31) Brown, A.; Rundqvist, S. Refinement of the Crystal Structure of Black Phosphorus. *Acta Crystallogr.* **1965**, *19*, 684–685 DOI:10.1107/S0365110X65004140.

(32) Yang, X. L.; Xie, S. W. Expression of Third-Order Effective Nonlinear Susceptibility for Third-Harmonic Generation in Crystals. *Appl. Opt.* **1995**, *34*, 6130–6135 DOI:10.1364/AO.34.006130.

(33) Keller, O. Random-Phase-Approximation Study of the Response Function Describing Optical Second-Harmonic Generation from a Metal Selvedge. *Phys. Rev. B* **1986**, *33*, 990–1009 DOI:10.1103/PhysRevB.33.990.

(34) Nemilentsau, A. M.; Slepyan, G. Y.; Khrutchinskii, A. A.; Maksimenko, S. A. Third-Order Optical Nonlinearity in Single-Wall Carbon Nanotubes. *Carbon N. Y.* **2006**, *44*, 2246–2253 DOI:10.1016/j.carbon.2006.02.035.

(35) Rudenko, A. N.; Yuan, S.; Katsnelson, M. I. Toward a Realistic Description of Multilayer Black Phosphorus: From G W Approximation to Large-Scale Tight-Binding Simulations. *Phys. Rev. B* **2015**, *92*, 85419 DOI:10.1103/PhysRevB.92.085419.

(36) Quereda, J.; San-Jose, P.; Parente, V.; Vaquero-Garzon, L.; Molina-Mendoza, A. J.; Agraït, N.; Rubio-Bollinger, G.; Guinea, F.; Roldán, R.; Castellanos-Gomez, A. Strong Modulation of Optical Properties in Black Phosphorus through Strain-Engineered Rippling. *Nano Lett.* **2016**, *16*, 2931–2937 DOI:10.1021/acs.nanolett.5b04670.

(37) Boyd, R. W. *Nonlinear Optics, Third Edition*; 3rd ed.; Academic Press: Burlington, Mass., 2008.

(38) Karvonen, L.; Säynätjoki, A.; Mehravar, S.; Rodriguez, R. D.; Hartmann, S.; Zahn, D. R. T.; Honkanen, S.; Norwood, R. A.; Peyghambarian, N.; Kieu, K.; Lipsanen, H.; Riikonen, J. Investigation of Second- and Third-Harmonic Generation in Few-Layer Gallium Selenide by Multiphoton Microscopy. *Sci. Rep.* **2015**, *5*, 10334 DOI:10.1038/srep10334.

(39) Mao, N.; Tang, J.; Xie, L.; Wu, J.; Han, B.; Lin, J.; Deng, S.; Ji, W.; Xu, H.; Liu, K.; Tong, L.; Zhang, J. Optical Anisotropy of Black Phosphorus in the Visible Regime. *J. Am. Chem. Soc.* **2016**, *138*, 300–305 DOI:10.1021/jacs.5b10685.

(40) Woodward, R. I.; Murray, R. T.; Phelan, C. F.; de Oliveira, R. E. P.; Runcorn, T. H.; Kelleher, E. J. R.; Li, S.; de Oliveira, E. C.; Fechine, G. J. M.; Eda, G.; de Matos, C. J. S. Characterization of the Second- and Third-Order Nonlinear Optical Susceptibilities of Monolayer MoS2 Using Multiphoton Microscopy. **2016**.

(41) Säynätjoki, A.; Karvonen, L.; Rostami, H.; Autere, A.; Mehravar, S.; Lombardo, A.; Norwood, R. A.; Hasan, T.; Peyghambarian, N.; Lipsanen, H.; Kieu, K.; Ferrari, A. C.; Polini, M.; Sun, Z. Ultra-Strong Nonlinear Optical Processes and Trigonal Warping in MoS2 Layers. **2016**.

(42) Säynätjoki, A.; Karvonen, L.; Riikonen, J.; Kim, W.; Mehravar, S.; Norwood, R. A.; Peyghambarian, N.; Lipsanen, H.; Kieu, K. Rapid Large-Area Multiphoton Microscopy for Characterization of Graphene. *ACS Nano* **2013**, *7*, 8441–8446 DOI:10.1021/nn4042909.

(43) Savostianova, N. A.; Mikhailov, S. A. Giant Enhancement of the Third Harmonic in Graphene Integrated in a Layered Structure. *Appl. Phys. Lett.* **2015**, *107* DOI:10.1063/1.4935041.

(44) Hendry, E.; Hale, P. J.; Moger, J.; Savchenko, A. K.; Mikhailov, S. A. Coherent Nonlinear Optical Response of Graphene. *Phys. Rev. Lett.* **2010**, *105* DOI:10.1103/PhysRevLett.105.097401.

(45) Cheng, J. L.; Vermeulen, N.; Sipe, J. E. Third Order Optical Nonlinearity of Graphene. *New J. Phys.* **2014**, *16* DOI:10.1088/1367-2630/16/5/053014.

(46) Rodrigues, M. J. L. F.; de Matos, C. J. S.; Ho, Y. W.; Peixoto, H.; de Oliveira, R. E. P.; Wu, H.-Y.; Neto, A. H. C.; Viana-Gomes, J. Resonantly Increased Optical Frequency Conversion in Atomically Thin Black Phosphorus. *Adv. Mater.* **2016** DOI:10.1002/adma.201603119.